\documentclass[11pt,letterpaper]{article}
\usepackage[margin=0.76in,headheight=14pt,headsep=16pt,footskip=24pt]{geometry}
\usepackage[T1]{fontenc}
\usepackage{lmodern,microtype,amsmath,amssymb,graphicx,caption,float,fancyhdr,booktabs,longtable,array}
\usepackage[hidelinks]{hyperref}
\hypersetup{pdftitle={Scaling Bayesian Bandit Encoding with Shared Learning},pdfauthor={Bhaskar Krishnamachari},pdfsubject={Large-catalog extension of the closed-loop Bayesian bandit encoder}}
\newcommand{\figdir}{figures}
\newcommand{\revfigdir}{figures}
\newcommand{\PrunedGain}{33.5}
\newcommand{\PrunedLo}{8.3}
\newcommand{\PrunedHi}{55.7}
\newcommand{\PrunedLatentRegret}{4.24}
\newcommand{\PrunedIndependentRegret}{6.37}
\newcommand{\StaticRegret}{6.81}
\newcommand{\SevereStaticRegret}{61.29}
\begin{document}
\thispagestyle{plain}
\begin{center}
{\LARGE\bfseries Scaling Bayesian Bandit Encoding\\[3pt] with Shared Learning\par}
\vspace{6pt}
{\normalsize Bhaskar Krishnamachari\\[3pt]
Ming Hsieh Department of Electrical and Computer Engineering\\
Viterbi School of Engineering, University of Southern California\\
\href{mailto:bkrishna@usc.edu}{bkrishna@usc.edu}}
\end{center}
\vspace{-6pt}
\begin{abstract}
A communication system must choose error protection and decoding effort as channel
conditions change. A Bayesian bandit encoder (BBE) uses receiver feedback to learn
which transmission configuration to select. We study a receiver that decodes by
guessing error patterns, using Guessing Random Additive Noise Decoding (GRAND).
We extend BBE's selection component to
1,008 code and decoder configurations by learning shared performance patterns
offline and updating their weights online. Decoder noise models remain fixed.
On a six-configuration training-selected
shortlist, sharing reduces accumulated utility loss by \PrunedGain\% relative to
independent learning. A fixed training-selected configuration matches the shared
learner that searches the full catalog. After channel changes, the pruned shared
learner first meets a near-optimal selection criterion in 88.5\% of events by
2,000 packets, compared with 54.2\% for pruned independent learning with the same
discounting. The results support combining sharing and pruning for configuration
selection, although packet losses remain high for the tested codes under severe noise.
\end{abstract}

\section{Introduction}
A wireless transmitter adds redundant bits to a packet so that the receiver can
correct errors introduced by the channel. The \emph{code rate} is the fraction of
transmitted bits carrying information. More redundancy can improve recovery but
reduces this fraction. The arrangement of errors also matters: scattered flips and
bursts can favor different codes and decoding strategies.

Guessing Random Additive Noise Decoding (GRAND) tries possible error patterns,
reverses their proposed flips, and tests whether the result is a valid encoded
packet~\cite{grand}. Each test is a \emph{query}. The code determines the validity
test, while a noise model determines the query order. A query limit bounds decoding
effort. An \emph{interleaver}, a bit permutation reversed at the receiver, provides
another choice by changing which code positions a burst affects.

Our starting point is the recently-proposed closed-loop Bayesian bandit encoder
(BBE)~\cite{bbe}. It couples Bayesian interference estimation with decoder
adaptation and selection between a random linear code and its interleaved version.
The learned interference parameters feed a hidden Markov model (HMM), which infers
bit-flip probabilities used to reorder GRAND queries. This decoder update changes
which transmission mode performs best. We build on that architecture by asking
how packet feedback can guide selection among many codes, interleavers, and decoder
settings.

We extend BBE's selection component. Each complete configuration is an \emph{arm};
the available set is the \emph{catalog}. Arms vary the code, within-codeword
interleaver, noise ordering, and query limit. We reward information delivery and
charge for queries. We fit decoder noise models offline and hold them fixed while
the selector learns arm performance online. This isolates configuration selection
within the broader BBE architecture.

A packet supplies feedback only for the selected arm. Thompson sampling addresses
this limited feedback by maintaining uncertainty about performance, sampling
plausible predictions, and selecting the largest predicted score~\cite{russo}.
Independent learning updates one arm at a time. We instead learn a \emph{latent
model}: a few shared, unobserved variables describe how many arms perform. Training
simulations supply a performance matrix. During transmission, one observation
updates the shared variables and hence predictions for other arms.

We also use \emph{pruning} to select a shortlist from training data. We measure
shared structure across 1,008 configurations, separate sharing from pruning,
compare with two fixed-arm baselines, and evaluate adaptation after channel changes.
Our contribution is this empirical study of a telemetry representation, a GRAND
configuration catalog, and their combination with pruning. Latent Thompson
sampling for link adaptation has a close precedent~\cite{saxena}.
Section~2 reviews related work, Section~3 describes the learner, and Sections~4--6
present results and conclusions. The appendix specifies the model and experimental procedures.

\section{Related Work}
\subsection{Noise-aware decoding}
Duffy, Li, and M\'edard introduced GRAND and its query-limited variant,
GRANDAB~\cite{grand}. For an input-independent additive-noise channel, GRAND
with the correct likelihood order finds a maximum-likelihood decoding.
An, M\'edard, and Duffy developed GRAND Markov Order
to exploit correlated errors and showed that keeping bursts can outperform
interleaving in suitable regimes~\cite{grandmo}. These results motivate treating
the error-pattern ordering and interleaver as parts of the selected configuration.

Soft GRAND uses received-signal confidence when ordering guesses~\cite{sgrand}.
Ordered Reliability Bits GRAND (ORBGRAND) uses reliability ranks to obtain an
ordering suited to efficient implementation~\cite{orbgrand}. Soft-output GRAND
also estimates confidence in the decoded result and supports iterative decoding
of longer codes~\cite{sogrand}. These methods offer ways to improve reliability
and enrich feedback beyond the hard-decision experiments reported here.

Willems, Shtarkov, and Tjalkens' context-tree weighting method combines predictions
from variable-memory binary sources~\cite{ctw}. Our context-based ordering uses
smoothed predictions from histories of different lengths, with shorter histories
receiving more weight when data are scarce. It is a practical backoff approximation
inspired by this approach. Miyamoto and Yang study noise guessing when the
finite-state channel law is unknown~\cite{universal}. Their focus is reliable
decoding without a known noise law; ours is learning the relative utility of
available code and decoder configurations.

\subsection{Sharing observations in link adaptation}
Combes et al. formulate transmission-rate selection as a bandit problem and
exploit structure among rate and mode choices~\cite{ors}. Saxena, Tullberg, and
Jald\'en provide a closer precedent for our learner~\cite{saxena}. Their latent
Thompson sampler updates a distribution over signal-to-interference-and-noise ratio
from packet acknowledgments. An offline link model maps that shared channel
estimate to success probabilities for all modulation and coding choices. Thus,
one packet teaches the learner about choices it did not make. Their arms are
modulation and coding schemes in an LTE link with Turbo coding. Their experiments
do not use GRAND receivers or select among receiver algorithms. They also evaluate time-varying
pedestrian and vehicular fading, using a Doppler-dependent spread of the SINR
posterior to track changes. Thus both studies address shared learning and changing
channels. Our study instead learns a multidimensional representation from decoder
telemetry, selects complete code--interleaver--GRAND configurations, and tests how
sharing and pruning interact across different temporal error patterns.

Joint-Thompson sampling uses another form of sharing: a joint prior that preserves
the ordering of success probabilities across modulation and coding
schemes~\cite{jointts}. We learn the relationships from a performance matrix
instead of imposing a single quality axis or an ordering across configurations.
This accommodates arms whose relative strengths depend on error arrangement as
well as error rate.

\subsection{Low-rank models, pruning, and changing channels}
Gopalan, Maillard, and Zaki study low-rank rewards generated by latent user
mixtures~\cite{gopalan}. Kveton et al. study efficient search for a large entry in
a stochastic low-rank matrix~\cite{kveton}. These models motivate learning shared
factors, while linear Thompson sampling provides the online inference
mechanism~\cite{agrawal}. In our setting, arm features are learned offline and
shared performance variables are inferred online.

Our shortlist uses the greedy set-cover principle~\cite{chvatal}: retain arms
that together perform well across training conditions. Pruning reduces the number
of choices; latent learning shares evidence among those choices. For time-varying
rewards, Qi, Wang, and Zhu analyze discounted Thompson sampling~\cite{discounted}.
We use discounting to track channel changes and evaluate its behavior under an
approximate, training-derived performance model.

\section{What the Learner Observes and Optimizes}
\subsection{Packet feedback and utility}
For packet $t$, let $a_t$ be the selected arm and $r_{a_t}$ its code rate. The
receiver reports success $S_t$ (one for correct recovery, zero otherwise),
abandonment $B_t$ (one if decoding gives up), and query count $Q_t$. We call these
three measurements \emph{telemetry}. The packet's score, or \emph{utility}, is
\begin{equation}
u_t=r_{a_t}S_t-\lambda Q_t,
\end{equation}
where $\lambda\geq0$ is the cost per query. Success earns the information
fraction transmitted, while queries incur a cost. Let $q_a$ be arm $a$'s query
limit and $Q_{\max}=\max_a q_a$ the largest limit in the catalog.
This is our chosen utility per decision. With unequal packet lengths, its average
is not aggregate delivered bits divided by aggregate transmitted bits. A lower-rate
code must recover enough additional packets to compensate for its redundancy.
The block error rate (BLER) is the probability
that a packet is not recovered correctly. For channel condition $z$, the mean
utility of arm $a$ is $U(a,z)=r_a[1-\mathrm{BLER}(a,z)]-\lambda\mathbb{E}[Q\mid a,z]$.

Only the selected arm's telemetry reaches the learner. Channel labels and outcomes
for unselected arms remain with the evaluator. Success comes from simulator ground
truth: accepting a valid codeword does not establish that it is the transmitted
word. The feedback therefore represents ideal error detection with no overhead.

\subsection{How the performance matrix enables sharing}
Suppose the catalog contains $A$ arms, training covers $Z$ channel conditions,
and each packet supplies $m$ telemetry measurements. Stacking each arm's mean
telemetry gives a matrix with $mA$ rows and $Z$ columns. We normalize queries by
$Q_{\max}$, subtract each arm's training-channel mean, and scale each measurement
type by its standard deviation across arm--channel entries. A positive scale
floor $s_{\min}$ prevents division by very small values. We fit shared patterns
using singular-value decomposition and select the retained number $d$, or
\emph{rank}, by prediction error on withheld training data.
Appendix~\ref{app:fit} gives the complete transformation and validation procedure.

Let $\mathbf y_t\in\mathbb R^m$ denote normalized packet feedback, and let
$\boldsymbol\theta$ contain the shared variables describing the current channel.
Training supplies a baseline prediction $\mathbf b_a$ for each arm and a matrix
$F_a\in\mathbb R^{m\times d}$ that maps the $d$ shared variables to changes in
predicted telemetry. Both $\mathbf b_a\in\mathbb R^m$ and $F_a$ are transformed
back from standardized coordinates, so the online equation uses the feedback
units defined above:
\begin{equation}
\mathbb{E}[\mathbf y_t\mid a_t,\boldsymbol\theta]
\approx \mathbf b_{a_t}+F_{a_t}\boldsymbol\theta.
\end{equation}
When an observation changes the estimate of $\boldsymbol\theta$, predictions for
every arm change through its corresponding $F_a$. The shared variables are
numerical features learned from performance data. They need not correspond directly
to bit error probability or burst length, or to one of the training channels.

The prior is $\boldsymbol\theta\sim\mathcal N(\mathbf0,I_d)$, where $I_d$ is the
identity matrix. We use a Gaussian observation model with an arm-specific
$m\times m$ covariance $R_a$, estimated from within-packet training telemetry.
It includes correlations among measurements. Without discounting, the
update gives the Gaussian posterior under the fixed-parameter linear observation
model. With discounting, it gives a weighted-likelihood Gaussian update that
exponentially downweights older observations while retaining the prior.
This is an approximation for bounded packet telemetry; its uncertainty estimates
have not been calibrated against measurements from a physical receiver.
Write utility as $\mathbf w_a^T\mathbf y_t$, where $\mathbf w_a$ contains the
measurement weights. For success, abandonment, and normalized queries,
$\mathbf y_t=(S_t,B_t,Q_t/Q_{\max})^T$ and
$\mathbf w_a=(r_a,0,-\kappa)^T$, with $\kappa=\lambda Q_{\max}$.
Abandonment informs inference but has no separate utility penalty.

\begin{figure}[H]
\centering\fbox{\begin{minipage}{0.94\linewidth}
\small\textbf{Algorithm 1: shared telemetry Thompson sampling.}
Given an available set, fixed $\mathbf b_a,F_a,R_a$, and discount $\gamma$, initialize
precision $P=I_d$ and information vector $\mathbf h=\mathbf0$.
For each packet:
\begin{enumerate}\setlength{\itemsep}{0pt}
\item Sample $\widetilde{\boldsymbol\theta}\sim\mathcal N(P^{-1}\mathbf h,P^{-1})$.
\item For each available arm, predict $\widetilde{\mathbf y}_a=\mathbf b_a+F_a\widetilde{\boldsymbol\theta}$
and score $\widetilde u_a=\mathbf w_a^T\widetilde{\mathbf y}_a$.
Clip this score to $[-\lambda q_a,r_a-\lambda]$, where $q_a$ is its query limit.
Select a maximizing arm, breaking ties uniformly at random.
\item Observe only that arm's packet feedback $\mathbf y$ and update
\[
P\leftarrow I_d+\gamma(P-I_d)+F_a^TR_a^{-1}F_a,\qquad
\mathbf h\leftarrow\gamma\mathbf h+F_a^TR_a^{-1}(\mathbf y-\mathbf b_a).
\]
\end{enumerate}
The next decision uses this posterior. Set $\gamma=1$ to retain all evidence,
or $0<\gamma<1$ to discount it. Decoder models stay fixed.
\end{minipage}}
\end{figure}

Only the sampled utility is clipped. Predicted telemetry and posterior updates
are not clipped. Independent Thompson sampling assigns each arm an independent
Gaussian mean with initial distribution
$\mathcal N(\mathbf b_a,F_aF_a^T)$, matching the shared model's marginal for that
arm. Both learners use the same $R_a$ and all three feedback measurements.
Independent discounting ages every arm's accumulated evidence at each update,
including unselected arms. Appendix~\ref{app:independent} specifies the updates.

\subsection{Choosing a shortlist or a fixed arm}
We prune using training utilities alone. An arm covers a training condition if its
mean utility is within a tolerance $\delta$ of the best arm for that condition. We repeatedly add
the arm that covers the most conditions still uncovered. This produces a shortlist
whose members work well under different training conditions. Independent pruned
and latent pruned use the same shortlist and keep their original priors. The full
versions can select any arm in the available catalog.

We consider two \emph{static baselines}, both of which keep one arm for every
test packet. The diverse-training baseline chooses the arm with the highest
average utility over the equally weighted training conditions. The severe-IID
baseline chooses the best arm for a single severe channel with independent,
identically distributed (IID) bit flips. It represents a non-adaptive decision
calibrated without exposure to diverse channel models. For this illustrative
comparator we use reference estimates for the noisiest evaluated IID condition,
so its selection is test-informed. The learners, their priors, the shortlist,
and the diverse-training static choice use training data only.

\section{Experiments I: Learning on a Fixed Channel}
We now specify the experimental values used for the general method above.
There are $A=1{,}008$ arms, $Z=24$ training conditions, and $m=3$ feedback
measurements: success, abandonment, and normalized query count. The resulting
training matrix is $3{,}024\times24$. Training validation selects $d=2$;
the scale floor is $s_{\min}=0.03$ and the pruning tolerance is $\delta=0.01$.
We set $Q_{\max}=16{,}384$ and $\lambda=10^{-6}$, so
$\kappa=\lambda Q_{\max}=0.016384$. This utility design gives decoding effort
a modest cost: using the maximum budget subtracts about 1.64 percentage points
from rate-weighted success. The value is a design choice, not an empirically
estimated channel parameter. Fixed-channel trials use $\gamma=1$; discounted
channel-change trials use $\gamma=0.99$.

We hold the channel fixed during each trial. The catalog combines 16 short codes
from Polar, Reed--Muller, random-linear, and low-density parity-check (LDPC) families.
Each carries 16 information bits in 24, 32, 40, or 48 transmitted bits. Combining
four interleavers, four GRAND orderings, and four query limits, then removing
duplicates, gives 1,008 arms. The orderings model independent flips, dependence on
the previous error bit, variable-length error histories, or burst durations.
We fit the noise models on training noise and hold them fixed during testing.
Table~\ref{tab:setup} summarizes the design. Appendix~\ref{app:catalog} identifies
every code construction and retained configuration.

\begin{table}[H]\centering\small
\caption{Experimental design. Catalog size counts configurations, not codewords.}
\label{tab:setup}
\begin{tabular}{ll}\toprule
Choice & Values \\\midrule
Information and transmitted lengths & $k=16$; $n=24,32,40,48$ \\
Interleavers & Identity, four-row transpose, two seeded permutations \\
GRAND query limits & 64; 512; 4,096; 16,384 \\
Available catalog sizes & Nested subsets of 32, 128, 512, 1,008 arms \\
Shared representation & 2 variables, fitted on all 1,008 training arms \\
Fixed-channel evaluation & 128 trials per method and size; 600 packets \\
Changing-channel evaluation & 48 trials per method; 600 + 2,000 + 2,000 packets \\\bottomrule
\end{tabular}\end{table}

The 24 training conditions include independent flips, isolated errors separated
by correct bits, correlated errors, bursts, slowly varying error probabilities,
and periodic interference. Their nominal bit error probabilities range from 0.015
to 0.12. The 16 test conditions use new parameters at probabilities 0.035 and 0.10,
including long-tailed burst durations and unequal probabilities for the two
directions of bit flips. We use 512 training packets per condition, a separate
1,024-packet collection for online sampling, and another independent 4,096 packets
to estimate each arm's mean utility. These counts apply to every arm--channel
pair. Outcomes share packet noise across arms, and the three collections use
independent random streams. Online trials sample packet indices with replacement
from the 1,024-packet collection, which is reused across trials. Channel changes
switch collections. Within-codeword error structure is preserved, but successive
packets do not continue the same noise trajectory. Thus slow variation and
periodicity describe structure within a packet. Appendix~\ref{app:sampling}
describes packet sharing and matching across methods.

We call the largest estimated mean the \emph{reference optimum}.
\emph{Cumulative regret} sums the difference between this optimum and the selected
arm's mean utility over a trial. It measures utility lost while learning. Even
pruned methods are compared with the best arm in the full available catalog.
Each method runs 128 matched trials of 600 packets per catalog size, balanced
across the 16 test conditions, observing only its selected arm's feedback.

\begin{figure}[H]
\centering\includegraphics[width=\linewidth]{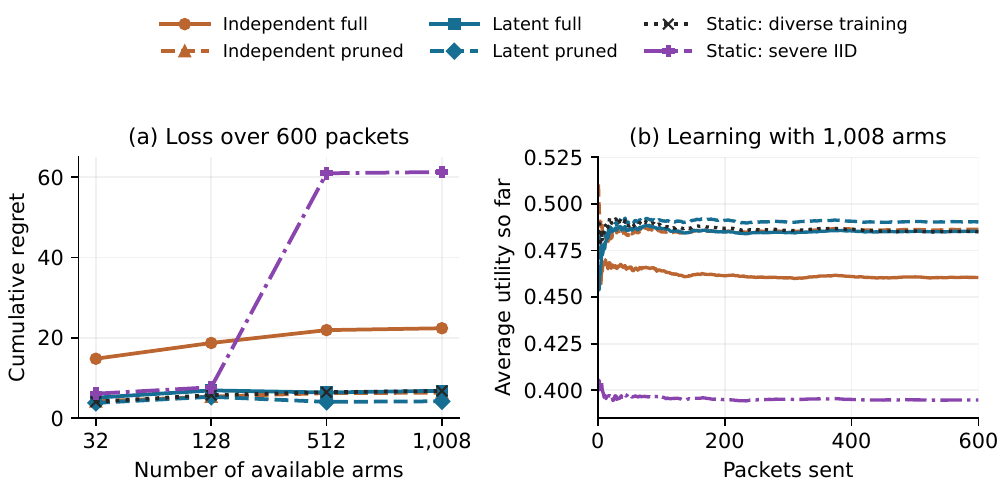}
\caption{\textbf{Sharing and pruning both reduce learning cost.} Full methods can
choose any available arm; pruned methods use the same training-selected shortlist
(3, 5, 6, and 6 arms as the catalog grows). Both static choices are fixed online:
one maximizes average training utility, while the other maximizes reference-estimated
utility at IID bit error probability 0.10 and is test-informed.
(a) Cumulative regret over 600 packets, lower is better. (b) Average observed
utility so far, higher is better. Each curve averages 128 trials.}
\label{fig:scaling}
\end{figure}

At 1,008 arms, latent full has cumulative regret 6.85, compared with 22.41 for
independent full and \StaticRegret{} for the static baseline
(Figure~\ref{fig:scaling}). A static choice based on diverse training thus matches
the full latent learner.
On the same six-arm shortlist, independent learning has regret
\PrunedIndependentRegret{} and latent learning has regret \PrunedLatentRegret.
Sharing reduces regret by \PrunedGain\% in this controlled comparison, with a
95\% bootstrap interval of \PrunedLo\% to \PrunedHi\%. We compute the interval
by resampling channels and then matched trials within channels.
The 1,000 bootstrap draws hold training, packet collections, and reference
estimates fixed, so this interval describes conditional trial variability.

The severe-IID static choice has regret \SevereStaticRegret{} at 1,008 arms,
with average observed utility about 0.395, compared with about 0.486 for the
diverse-training static choice. Its regret rises sharply when the catalog grows
from 128 to 512 arms because a rate-$1/2$ Reed--Muller configuration becomes
available and wins the severe-IID comparison. That choice sacrifices utility
on other channels relative to the previously selected rate-$2/3$ random-linear
configuration. This illustrates the risk of deploying a fixed configuration
chosen for one channel model. It does not establish that learning always beats
a well-calibrated static choice: the diverse-training baseline remains strong.

The rank-two telemetry model has root-mean-square utility prediction error 0.0247
on withheld training measurements. A separate diagnostic removes arm and channel
averages from the $1,008\times24$ scalar utility matrix. Ranks 3, 4, and 7 explain
90\%, 95\%, and 99\% of its squared variation. This utility diagnostic and the
prediction-based telemetry rank selection concern different matrices and criteria.

For each of the 16 test conditions, we also identify the arm maximizing reference
mean utility. These utility-optimal arms have BLER from 7.9\% to 34.9\%, rather
than being minimum-BLER choices. Efficient selection in this catalog therefore
still leaves substantial packet loss. Training represents 12,386,304 arm--packet
outcomes. Recorded training-bank processing took 10.61 seconds using batched
syndrome lookups, and the analysis stage took 2.80 seconds.
These partial timings exclude some setup work and are not receiver
latencies or an end-to-end training benchmark (Appendix~\ref{app:cost}).

\section{Experiments II: Learning after Channel Changes}
We test 12 ordered channel-change scenarios involving 11 distinct channel
conditions. Training predictions select four changes in error probability, four
in error arrangement, and four into a family absent from training. Each scenario
runs A for 600 packets, B for 2,000, then A again for 2,000. Four repetitions give
48 trials and 96 transitions. The learner discounts old observations by 0.99 at
every update while retaining its prior, without change notifications or resets.

\begin{figure}[H]
\centering\includegraphics[width=\linewidth]{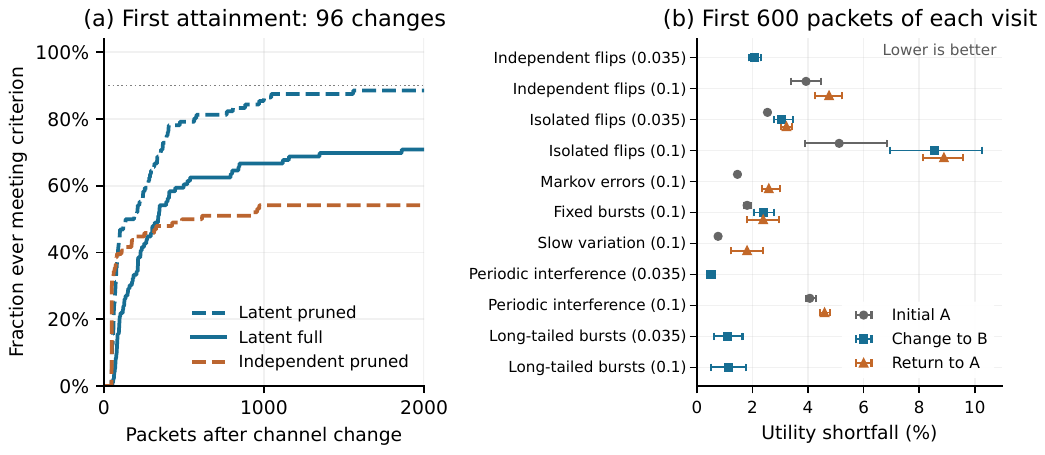}
\caption{\textbf{First attainment and performance by destination.} (a) All three
learners discount by 0.99; the dotted line marks 90\% of events. (b) Six-arm pruned
latent learner; parentheses give nominal bit error probability. Markers average
4--16 visits over 1--4 scenarios; bars are 95\% bootstrap intervals, resampling
scenarios and then trials, with packet banks and reference estimates fixed.
Missing markers indicate untested roles. For a single scenario, the interval
reflects repetition variability only.}
\label{fig:recovery}
\end{figure}

Let $U^*(z)=\max_a U(a,z)$ denote the full-catalog reference optimum, which lies
between 0.3198 and 0.6142 across all 16 test conditions. For local packet index $t$
after a change to $z$, Figure~\ref{fig:recovery}(a) records the first attainment time
\begin{equation}
\tau=\inf\left\{t\geq50:\sum_{j=t-49}^{t}
\mathbf1\{U(a_j,z)\geq0.95U^*(z)\}\geq45\right\}.
\end{equation}
The window contains only post-change packets. All changes are included, even
when the pre-change arm already meets the destination criterion. Such events can
first qualify at packet 50. An event stays counted if performance later falls,
so this curve is not the fraction currently maintaining the criterion.
Panel (b) shows \emph{utility shortfall}: the percentage
by which the selected arms' mean utility, averaged over the first 600 packets of
each visit, falls below the reference optimum. A 5\% shortfall means achieving
95\% of the best available mean utility. Its 11 rows group observations by
destination channel, with separate markers for initial A, change to B, and return
to A. A destination may occur in several scenarios.

Sharing and pruning help the selector find a suitable arm after a channel change.
Within 2,000 packets, the pruned latent learner meets the criterion in 85 of 96
changes (88.5\%), compared with 68 (70.8\%) for full latent and 52 (54.2\%)
for pruned independent learning. Keeping all past evidence reduces the pruned
latent learner's count to 60 (62.5\%). Discounting therefore helps in these trials.

\begin{table}[H]\centering\small
\caption{First attainment and sustained performance over all 96 changes.
The median retains unattained events in its denominator. The final column is the
mean fraction of near-optimal choices over packets 1,501--2,000 after each change.}
\label{tab:sustained}\begin{tabular}{lrrr}\hline
Learner & Attained / 96 & Median packet & Final 500 (\%) \\\hline
Latent pruned & 85 & 136 & 83.3 \\
Latent full & 68 & 339 & 62.8 \\
Independent pruned & 52 & 488 & 59.4 \\
Latent pruned, no discount & 60 & 1045 & 60.0 \\
\hline\end{tabular}
\end{table}

The pruned latent learner reaches the criterion sooner and uses good arms more
often afterward. Half of the changes have met the criterion by packet 136,
compared with packet 488 for pruned independent learning. Over the final 500
packets after each change, their fractions of near-optimal choices are 83.3\%
and 59.4\%, respectively (Table~\ref{tab:sustained}). Some changes remain
difficult: no method reaches the criterion in 90\% of changes within the
observation period. Every destination has an arm in the shortlist achieving at
least 98.6\% of the reference optimum, so a sufficiently good choice is available
even in the cases where the learner does not find or maintain it.

Figure~\ref{fig:recovery}(b) identifies which destination channels remain difficult.
For severe isolated flips, average utility falls 8.5\% below the optimum on
visits to B and 8.9\% on returns to A, compared with 5.1\% on initial visits.
The difficulty therefore appears on either type of transition to this channel.
For IID flips at probability 0.10, the shortfall is 3.9\% initially and 4.8\%
on return. These comparisons use the first 600 packets of every visit,
so longer return visits do not change the averaging interval. However, visits
to B can follow different source channels than returns to A. The figure locates
poor performance by destination; a separate experiment with matched histories
would be needed to measure a penalty caused specifically by returning.

\section{Conclusion and Limitations}
Latent sharing extends BBE's selector to a large catalog and improves selection
within a shortlist, with a 33.5\% regret reduction against independent learning on
the same six arms. Discounted shared learning also improves first attainment and
sustained near-optimal selection after changes. The tested utility-optimal arms
still incur substantial packet loss, and the offline model uses measurements of
every configuration. Future work will jointly learn the channel, adapt the decoder,
and learn which code--decoder configuration to select, combining BBE's adaptive
receiver with latent sharing and pruning.

\section*{AI Use Statement}
The author has used ChatGPT and Codex (primarily gpt-6) for writing code and
simulations, for help with algorithm design, for paper writing, editing, proofreading and formatting. The human author accepts full
responsibility for the contents of this work.

\clearpage\appendix
\section*{Appendices}
\section{Fitting the shared representation}\label{app:fit}
Let $Y_{azj}$ be training mean telemetry for arm $a$, channel $z$, and measurement
$j\in\{1,2,3\}$. The measurements are success, abandonment, and queries divided
by 16,384. There are $A=1,008$ arms and $Z=24$ channels. An observation mask
$O_{az}$ applies to all three measurements together. Let $n_a=\sum_zO_{az}$.
The initial offset $b^0_{aj}$ is the mean of $Y_{azj}$ over observed channels for
that arm. If $n_a=0$, we use the measurement's mean over all observed arm--channel
entries. The scale $s_j$ is its population standard deviation over those entries,
with a minimum of 0.03. Thus constant measurements remain well defined.

We arrange the centered measurements in a $3A\times Z$ matrix with row
$3a+(j-1)$ for arm index $a\in\{0,\ldots,A-1\}$ and measurement index
$j\in\{1,2,3\}$, counting matrix rows from zero:
\begin{equation}
X_{(a,j),z}=(Y_{azj}-b^0_{aj})/s_j.
\end{equation}
Unobserved entries initially equal zero. Each completion iteration takes a rank-$d$
SVD approximation and replaces only unobserved entries with their fitted values,
leaving observed entries unchanged. We run 20 iterations and then take a final
SVD. With a fully observed matrix, no completion is needed.

Write the final leading singular vectors and values as $L_d,D_d,V_d$. The online
features and training coordinates are initially
\begin{equation}
F_{aj,:}=s_j(L_dD_d)_{(a,j),:}/\sqrt{Z-1},\qquad
\Theta=\sqrt{Z-1}\,V_d.
\end{equation}
We subtract the row mean $\bar\theta$ from $\Theta$ and add $F_a\bar\theta$ to
$\mathbf b_a^0$. This preserves fitted means and centers the training coordinates.
The resulting $\mathbf b_a$ and $F_a$ are in the original feedback units, not
standardized units. We use the deployment prior $\mathcal N(0,I_d)$; this is a
modeling choice with scale set by the decomposition. For the fully observed fit,
the centered coordinates have sample covariance $I_d$, up to numerical precision.
The prior describes shared performance variation, not uncertainty in fitted
code or decoder parameters.

\subsection{Validation and rank choice}
A physical group consists of one code, interleaver, and noise ordering with four
query limits. For each of three random splits, we observe a physical-group--channel
entry with probability 0.8. All query-limit variants and all three measurements
share that mask. The implementation ensures at least one observed group per
channel. Transformations and completion use observed entries only.

We compare $d\in\{2,4,6,8\}$ by the root-mean-square utility error on withheld
arm--channel entries. We choose the smallest rank whose mean error is within one
standard error of the lowest mean error across the three splits. Mean errors are
0.02470, 0.04313, 0.06608, and 0.08522 respectively, selecting $d=2$.
Rank one was not a candidate. We then refit on all training entries. The smaller
catalog experiments use subsets of this full-catalog model, not independently
refitted representations. Both learners receive the corresponding same-arm priors.

The residual utility diagnostic instead forms $M_{az}=U(a,z)$ from training means
and analyzes $M_{az}-\bar M_{a\cdot}-\bar M_{\cdot z}+\bar M$ by SVD.
The 90\%, 95\%, and 99\% ranks refer to cumulative squared singular values of
this $1,008\times24$ matrix. They do not determine the telemetry model's rank.

\subsection{Observation covariance and inference}\label{app:independent}
For each arm and training channel, we calculate the unbiased $3\times3$ sample
covariance of 512 packet telemetry vectors, using denominator 511. We average
these matrices equally across the 24 channels and floor their eigenvalues at
$10^{-4}$ to obtain $R_a$. The three components can be correlated, but $R_a$ is
fixed across online channel conditions. This is within-packet variability, not
matrix reconstruction error. The implementation also applies a $10^{-7}$
positive-definiteness floor when loading the already regularized matrices.

Algorithm 1 uses precision $P$ and information vector $\mathbf h$, giving mean
$P^{-1}\mathbf h$ and covariance $P^{-1}$. Cholesky solves implement sampling and
updating. The discount equation $I_d+\gamma(P-I_d)$ retains the identity prior
while reducing accumulated evidence. For fixed-channel trials feedback is applied
after selection; in changing-channel trials it is queued and applied before the
next decision. With immediate feedback these give the same decision timing.

For independent learning, let the unknown mean telemetry of arm $a$ have prior
$\mathbf m_a\sim\mathcal N(\mathbf b_a,C_a)$, where $C_a=F_aF_a^T$.
Different arms are independent. The following specifies the implemented update
even when $C_a$ is singular. Diagonalize the whitened prior:
\begin{equation}
R_a^{-1/2}C_aR_a^{-1/2}=V_a\operatorname{diag}(e_{aj})V_a^T,
\quad T_a=V_a^TR_a^{-1/2},\quad D_a=R_a^{1/2}V_a.
\end{equation}
Negative eigenvalues from numerical roundoff are set to zero. Initially each arm
has count $c_a=0$ and information vector $\mathbf h_a=0$. Before incorporating
each observed packet, multiply \emph{all} $c_a$ and $\mathbf h_a$ by $\gamma$.
For the selected arm only, add one to $c_a$ and add
$T_a(\mathbf y-\mathbf b_a)$ to $\mathbf h_a$. Its whitened coordinate variance
and mean are
\begin{equation}
v_{aj}=\frac{e_{aj}}{1+c_ae_{aj}},\qquad
\widehat x_{aj}=v_{aj}h_{aj}.
\end{equation}
These expressions retain the prior and leave zero-variance directions fixed.
With $\mathbf w_a=(r_a,0,-\kappa)^T$, the implementation independently samples
each arm's utility from a normal distribution with mean
$\mathbf w_a^T(\mathbf b_a+D_a\widehat{\mathbf x}_a)$ and variance
$\sum_j[(\mathbf w_a^TD_a)_j]^2v_{aj}$. It uses the same utility clipping and
random tie-breaking as the shared learner. Clipping affects selection only.

\section{Codes and selectable configurations}\label{app:catalog}
Table~\ref{tab:codes} lists the 16 constructions. All carry $k=16$ information
bits. They are short representative binary linear codes constructed using the
procedures below. Randomized constructions are fixed before channel evaluation.

\begin{table}[H]\centering\small
\caption{All codes, with $k=16$. Mother length applies to Polar and Reed--Muller
constructions; the last column counts constant-zero transmitted positions.}
\label{tab:codes}
\begin{tabular}{lrrr}\toprule
Family & $n$ & Mother length & Zero positions \\\midrule
Polar & 24 & 32 & 0\\
Polar & 32 & 32 & 0\\
Polar & 40 & 64 & 0\\
Polar & 48 & 64 & 0\\
Reed--Muller & 24 & 32 & 0\\
Reed--Muller & 32 & 32 & 0\\
Reed--Muller & 40 & 64 & 0\\
Reed--Muller & 48 & 64 & 0\\
Random linear & 24 & -- & 0\\
Random linear & 32 & -- & 0\\
Random linear & 40 & -- & 0\\
Random linear & 48 & -- & 0\\
LDPC & 24 & -- & 0\\
LDPC & 32 & -- & 1\\
LDPC & 40 & -- & 1\\
LDPC & 48 & -- & 4\\\bottomrule
\end{tabular}\end{table}

For Polar codes, let $N=2^{\lceil\log_2 n\rceil}$ and $m=\log_2N$.
Form $\left[\begin{smallmatrix}1&0\\1&1\end{smallmatrix}\right]^{\otimes m}$.
Starting from erasure probability 0.5, recursively replace each reliability value
$x$ with $(2x-x^2,x^2)$. Select the 16 transform rows with the smallest final
values, using stable index order for ties. For Reed--Muller variants, evaluate
Boolean monomials on the $N$ binary points in lexicographic order. Select the
first 16 monomials, ordered by degree and then lexicographic variable combinations.
For $N=32$ these form the second-order Reed--Muller code; at $N=64$ they form a
16-dimensional subcode of the second-order code.

When $n<N$, consider 16 puncturing patterns: first the columns
$\lfloor\operatorname{linspace}(0,N-1,n)\rfloor$, then 15 sorted random
$n$-column subsets. Keep only full-rank punctured generators. At $n=N$ there is
one candidate. This is puncturing, with no payload shortening. Random-linear
construction considers eight generators $[I_{16}\ P]$ with independent fair bits
in $P$. LDPC construction tries up to 160 parity-check matrices of size
$(n-16)\times n$, each with three distinct uniformly selected checks per column.
It retains the first eight full-row-rank matrices, or fewer if the limit is reached,
and uses their binary nullspaces as generators.

Within each family and length, candidates are scored using the minimum weight
among 4,096 sampled messages (discarding all-zero messages) and all generator rows.
Candidates of a given family and length use the same sampled messages.
The largest score wins, with the earliest candidate breaking ties. This is a
sampled-weight screen, not a minimum-distance calculation. All ranks, nullspaces,
and canonical row reductions are computed over $\mathrm{GF}(2)$.

\subsection{Interleavers, budgets, and duplicate removal}
For a codeword $x$, the transmitted vector is $x[\pi]$. Both candidate noise and
received words use $\pi^{-1}$ before the same base-code membership test.
The four permutations are identity; \texttt{block4}, obtained by reshaping
$0,\ldots,n-1$ into four rows and flattening its transpose; and \texttt{random1}
and \texttt{random2}, two fixed random permutations per length.
Each applies within one codeword and adds no modeled
cross-packet buffering delay. Query limits are 64, 512, 4,096, and 16,384.

We hash the binary reduced-row-echelon generator after permutation. A duplicate
has the same $n,k$, canonical transmitted codebook, and decoder ordering. We
retain the first occurrence before adding query-limit variants. For the length-32
Reed--Muller code, \texttt{block4} is equivalent to identity. Removing its four
orderings at four budgets reduces 1,024 nominal configurations to 1,008, organized
in 252 physical groups.

The smaller catalogs are nested prefixes of a randomized balanced ordering, independent
of outcomes. We group the catalog by family, decoder ordering, and code length, shuffle indices
within groups, shuffle the groups, then repeatedly take one index from each group.
Query budget is not a balancing variable.

\begin{table}[H]\centering\small
\caption{The six retained configurations at full catalog size. All have $k=16$.
Budgets are shown separately because tie-breaking produces different query-limit
variants in the two experiments. The first row is also the stationary static arm.}
\label{tab:shortlist}
\begin{tabular}{lrllrr}\toprule
Code family & $n$ & Interleaver & Ordering & Fixed & Changing \\\midrule
Random linear &24& random1 & Context &4096&4096\\
Reed--Muller &32& random1 & IID &16384&16384\\
Random linear &24& identity & Context &4096&4096\\
Reed--Muller &24& block4 & Run length &16384&4096\\
Random linear &24& random2 & Context &16384&4096\\
Random linear &24& random2 & IID &16384&4096\\\bottomrule
\end{tabular}\end{table}

Pruning covers training conditions within absolute utility 0.01 of their best arm.
At each step, ties in coverage count are resolved by higher mean training utility,
then earlier catalog index. Selection stops when every condition is covered or
16 arms are retained. The fixed-channel experiment uses the balanced catalog
order; the channel-change experiment uses the original construction order.
Equal training utilities of query-limit variants explain the budget differences
in Table~\ref{tab:shortlist}. Within each experiment, the independent and shared
pruned learners receive exactly the same shortlist. It retains random-linear and
Reed--Muller codes, three noise orderings, and multiple permutations; the results
do not establish that the excluded families are intrinsically inferior.
At sizes 32 and 128 the static arm is random-linear $(24,16)$ with \texttt{random1},
respectively Markov and run-length ordering, and budget 512. At sizes 512 and 1,008
it is the first row of Table~\ref{tab:shortlist}.
The severe-IID static comparator instead selects random-linear $(24,16)$ with
identity interleaving and budget 4,096 at size 32; random-linear $(24,16)$ with
\texttt{random2} and budget 16,384 at size 128; and Reed--Muller $(32,16)$ with
identity and budget 16,384 at sizes 512 and 1,008. All four use IID ordering.
Selection maximizes reference-estimated utility at IID bit error probability
0.10, then freezes the arm across all test conditions.

\section{Frozen GRAND noise orderings}\label{app:noise}
We pool 64 independently generated 48-bit sequences from each of the 24 training
channels, totaling 1,536 sequences and 73,728 bits, generated independently of
the performance-estimation packets. We fit one model of each type on
this common pool and use it at all four lengths. Counts reset at sequence boundaries.

\paragraph{IID.} With $n_1$ observed ones among $N$ bits, the fitted probability
is $(n_1+1/2)/(N+1)=0.06726661$. Noise patterns are generated by increasing
Hamming weight; within a weight, bit-position combinations are lexicographic.

\paragraph{Markov.} We estimate the initial-bit probability and the two transition
rows using half-count smoothing. The resulting initial probability of one is
0.06148341, with $P(1\mid0)=0.04967842$ and $P(1\mid1)=0.31235886$.
Sequence probability is the initial probability times the product of transitions.

\paragraph{Context backoff.} We store next-bit counts for every observed suffix
of lengths zero through four. For context $c$ with counts $(n_{c0},n_{c1})$ and
$n_c=n_{c0}+n_{c1}$, its local estimate is
$q_c(b)=(n_{cb}+1/2)/(n_c+1)$. Start with the root estimate, then visit available
suffixes from shortest to longest and replace prediction $q$ by
\begin{equation}
q\leftarrow\frac{n_c}{n_c+2}q_c+\frac{2}{n_c+2}q.
\end{equation}
Unobserved suffixes leave the prediction unchanged. Probabilities are floored at
$10^{-12}$ and renormalized. The model combines several history lengths with
data-dependent weights. It is not the exact context-tree weighting mixture.

\paragraph{Run length.} The state is the last bit and its current run length,
capped at eight. Training counts switches and continuations for each state.
The switch probability is $(n_{\rm switch}+1/2)/(n_{\rm switch}+n_{\rm stay}+1)$,
with initial one-probability 0.06148341. Runs of length eight or more share a bin.
These are runs of observed error bits, not latent interference-burst durations.

For the three memory models, candidate generation uses a priority queue with
negative log prefix probability plus the minimum remaining suffix cost. Dynamic
programming computes that cost on the model's finite memory state. Complete
patterns are emitted in decreasing model probability; insertion order resolves
ties after the heap's cost keys. The limit is 1,000,000 expanded prefixes and
16,384 emitted patterns. Every length--ordering list reached 16,384
patterns, so the node limit did not truncate this study. This is exact ordering
under each fitted model, which may be mismatched to the actual channel.

We compute the syndrome of each inverse-permuted candidate once. A lookup retains
the earliest candidate for each syndrome. For a received packet, the first matching
candidate within the arm's budget is the GRAND decision; its one-based index is
the query count. If no candidate fits the budget, queries equal the budget and
decoding abandons. This lookup reproduces query-limited GRAND decisions without
timing a sequential decoder for every packet.

\section{Channel processes and parameter grids}\label{app:channels}
Let $p$ denote nominal marginal bit-error probability. Table~\ref{tab:channels}
specifies all conditions as Cartesian products of probability sets and the listed
family parameters: 24 training conditions and 16 test conditions. The processes
below define the corresponding transition and arrival probabilities.

\begin{table}[H]\centering\small
\caption{Complete training and test grids. Training uses
$p\in\{0.015,0.04,0.08,0.12\}$ in each of its six families. Testing uses
$p\in\{0.035,0.10\}$ in each of eight families. A dash means absent from training.}
\label{tab:channels}
\begin{tabular}{lll}\toprule
Family & Training parameter & Test parameter \\\midrule
Independent & Bernoulli($p$) & Bernoulli($p$)\\
Isolated & Forced gap $g=3$ & $g=2$\\
Markov & Persistence $\rho=0.6$ & $\rho=0.8$\\
Fixed-duration burst & Duration $B=8$ & $B=10$\\
Slow variation & Coherence length $L=8$ & $L=12$\\
Periodic & Period $T=12$, depth 0.95 & $T=14$, depth 0.95\\
Long-tailed burst & -- & Exponent 1.7, maximum 64\\
Asymmetric & -- & Crossover ratio 4\\\bottomrule
\end{tabular}\end{table}

Independent noise flips each bit with probability $p$. Isolated noise forces $g$
correct bits after a flip; outside this refractory interval, a flip occurs with
probability $p/(1-pg)$. Each sample discards $\max(100,20g)$ initial symbols.
Markov noise starts with a Bernoulli($p$) error bit and uses
$P(1\mid0)=p(1-\rho)$ and $P(1\mid1)=p+\rho(1-p)$, giving stationary error
probability $p$.

The two burst families use background flip probability $\ell=0.002$ and burst
flip probability $h=0.55$. Set occupancy $q=(p-\ell)/(h-\ell)$. On an off symbol,
initiate a burst on the following symbol with probability
$q/[\mathbb E[D](1-q)]$; no burst begins during an existing burst, and at least
one off symbol separates bursts. Fixed bursts have $D=B$ and discard
$\max(100,20B)$ initial symbols. Long-tailed bursts use
$P(D=d)=d^{-1.7}/\sum_{i=1}^{64}i^{-1.7}$ for $1\leq d\leq64$, with
$\mathbb E[D]=4.47960455$ and burn-in $\max(200,\lfloor30\mathbb E[D]\rfloor)$.

Slow variation selects probability $h=0.55$ with probability
$(p-0.002)/(0.55-0.002)$ independently for each $L$-symbol interval, and otherwise
uses 0.002. A uniform phase in $\{0,\ldots,L-1\}$ offsets the interval boundaries.
Periodic noise has conditionally independent flips with probability
$p+0.95p\cos(2\pi(i+\phi)/T)$, where each packet draws a new uniform integer
phase $\phi\in\{0,\ldots,T-1\}$. All these models reset when generating a packet.

Asymmetric noise has $P(1\mid\text{sent }0)=1.6p$ and
$P(0\mid\text{sent }1)=0.4p$. We generate independent uniform 16-bit messages,
encode and interleave them, then apply these input-dependent probabilities using
shared uniform random numbers across configurations. Every nonzero generator
column gives a fair transmitted-bit marginal. Constant-zero columns instead
always use $1.6p$. The LDPC codes in Table~\ref{tab:codes} with $z_0$ such columns
therefore have average error probability $p(1+0.6z_0/n)$, at most $1.05p$ here.
The nominal $p$ refers to a balanced input. We do not generate asymmetric noise
using an all-zero codeword. For input-independent channels, linear translation
invariance allows decoding the noise directly without generating a codeword.

An additional validation collection, included only in the offline-cost accounting
and excluded from the reported policy comparisons, has
12 conditions: the six training families at $p=0.025,0.06$, using gap 4,
Markov persistence 0.45, burst and coherence lengths 6, and period 10.

\section{Packet sampling and evaluation}\label{app:sampling}
The training, validation, test-replay, and test-reference collections use
independent random streams for each channel condition. Each packet
contains a newly generated 48-bit wire-noise sequence, whose prefixes serve
shorter codes. All arms share this noise, except that asymmetric noise is derived
from common random messages and uniforms as specified above. Query-limit variants
share the same decoder trace. Thus the 512, 512, 1,024, and 4,096 observations
per arm--channel pair are correlated across arms, not separately simulated samples.
The split streams are independent.

Each online step samples a column of the corresponding test-replay bank uniformly
with replacement. Only the selected arm's success, abandonment, and query count
from that column enter the learner. The bank is reused across trials. A changing
trial of 4,600 steps samples 600, 2,000, and 2,000 indices from the A, B, and A
collections, respectively, with separate random streams for each segment. Initial
and return visits do not receive identical packet sequences.

Fixed-channel trials are balanced across the 16 test conditions. Corresponding
trials use the same packet indices across methods, including both static baselines,
while posterior-sampling streams are method-specific. Changing-channel trials
likewise share packet indices across methods and use separate streams for each
segment. Matching the packet samples reduces variation in policy comparisons
without revealing any unselected arm's feedback to a learner.

Bootstrap intervals for the stationary regret reduction resample the 16 test
channels, then paired repetitions within each sampled channel (1,000 draws).
Figure 2(b) resamples scenarios within each destination and visit role, then
repetitions within each selected scenario (4,000 draws).
These intervals condition on the fitted model, cached packet banks, and estimated
reference utilities. They omit uncertainty from regenerating those objects.
The attainment and sustained-performance table reports descriptive averages,
without additional confidence intervals.

Attainment windows restart at each change. Unattained events are right-censored
at 2,000 packets and remain in the denominator for attainment percentiles.
The reported median is the earliest packet by which at least 48 of 96 events
have attained. The 90th percentile exceeds the observation horizon for all four
reported policies. Sustained performance averages the near-optimal indicator over
the final 500 packets in every event, including unattained events.

\subsection{Recorded offline cost}\label{app:cost}
There are 12,386,304 training arm--packet outcomes, representing 3,096,576
code--interleaver--ordering packet decodes shared across four query limits.
The training, validation, replay, and reference collections together represent
101,154,816 arm--packet outcomes. The saved per-channel processing times sum to
10.61, 7.86, 15.95, and 67.77 seconds, respectively. All records mark the bank as
newly generated, not loaded from cache. These timers include packet generation,
batched lookup, summaries, bank writing, and diagnostics, but exclude code
construction, model calibration, candidate preparation, and syndrome-lookup setup.

The 16 candidate-list preparation timers sum to 13.32 seconds. Their routine can
either generate a list or load an existing list, and its timing log does not
record which occurred. The saved analysis-stage timer is 2.80 seconds and includes
ancillary representation analyses. No complete first-run wall-clock time or peak
memory measurement for the reported physical pipeline is recorded. These timings
describe this batched simulator, not sequential GRAND latency or hardware throughput.

\subsection{Validation checks}
We tested posterior updates and information boundaries. Altering evaluator-only
reference values and unselected packet outcomes left the learner's choices
unchanged. For small blocks, we checked candidate ordering against exhaustive
likelihood sorting.

\end{document}